\documentclass[11pt]{article}
\usepackage{fullpage}
\title{The Complexity of Diagonalization}
\author{Nikhil Srivastava\footnote{UC Berkeley, \texttt{nikhil@math.berkeley.edu}. Supported by NSF Grant CCF-2009011. Invited survey for ISSAC 2023. }}
\date{\today}
\usepackage{amsmath,amsthm}
\newtheorem{theorem}{Theorem}[section]
\newcommand{\C}{\mathbb{C}}
\newcommand{\ax}[1]{\tilde{#1}}
\newcommand{\emach}{\mathbf{u}_{\mathrm{mach}}}
\newcommand{\fl}{\mathsf{fl}}
\newcommand{\kappaeig}{\kappa_\mathrm{eig}}

\theoremstyle{definition}
\newtheorem{remark}[theorem]{\bf Remark}

\newcommand{\bit}[1]{\langle #1\rangle}
\newcommand{\poly}{\mathrm{poly}}

\usepackage[pagebackref]{hyperref}

\usepackage[dvipsnames]{xcolor}
\hypersetup{
pagebackref,
letterpaper=true,
colorlinks=true,
urlcolor=blue,
linkcolor=blue,
citecolor=OliveGreen,
}
\usepackage{mathpazo}

\begin{document}

\maketitle

\begin{abstract}
    We survey recent progress on efficient algorithms for approximately diagonalizing a square complex matrix in the models of rational (variable precision) and finite (floating point) arithmetic. This question has been studied across several research communities for decades, but many mysteries remain. We present several open problems which we hope will be of broad interest.
\end{abstract}
\section{Introduction}

We  survey what is known and discuss what remains to be discovered about the complexity of the following computational problem, which is fundamental throughout mathematics, science, and engineering: 
\begin{quote} Given a diagonalizable $n\times n$ real or complex matrix $A$, {\em approximately compute} a diagonal matrix $D$ (containing the eigenvalues of $A$) and an invertible matrix $V$ with columns of unit $\ell_2$ norm (containing the right eigenvectors of $A$) such that $A=VDV^{-1}$.
\end{quote}
We  refer to such a factorization as an {\em exact diagonalization} of $A$, and to a $V$ with columns of unit $\ell_2$ norm as {\em normalized}; note that the exact diagonalization is unique only if the eigenvalues of $A$ are simple. This problem has been studied from various perspectives in at least four research communities --- theoretical computer science, symbolic computation, numerical analysis, and real number computation. The goal of this purely expository article is to give a unifying view of the research frontier across all of these communities and highlight some directions for future work. The choice of topics is is biased by the theoretical computer science background of the author: we will be concerned with provable (as opposed to heuristic or empirical) bounds on worst case complexity in this survey, where the input is assumed to be {\em arbitrary} (i.e., not necessarily symmetric or Hermitian, though we will occasionally comment on those important cases). 

There are many ways to formalize what is meant by  {\em complexity, compute,} and {\em approximate}. 
The measure of complexity depends on the choice of computational model, which will be one of the following. 
\begin{enumerate}
\item [(0)] {\bf Exact Real Arithmetic.} Numbers are stored exactly as real and complex numbers. The  complexity of an algorithm in this model is simply the total number of arithmetic operations, often referred to as {\em arithmetic complexity}. This is an idealized model that does not correspond to any actual computing devices. It is popular in the theoretical computer science and pure mathematics (real number compuation) communities, and often serves as a first setting to consider before moving on to more demanding models.
\item [(1)] {\bf Variable Precision Rational Arithmetic.} Numbers are stored as fractions $p/q$,
which require $\lg_2(|p|)+\lg_2(|q|)$ bits to store and nearly linear time to perform
exact addition, multiplication, and division with. The bit lengths of numbers may grow under arithmetic operations (in particular, multiplication can double the bit length). The complexity of the algorithm must 
take the cost of arithmetic into account; in particular, if the algorithm uses  integers of length at most $B$, then the bit complexity is at most the arithmetic complexity times\footnote{Throughout this article, the $O^*$ notation suppresses logarithmic factors in the input size parameters $n,a$ and $\log\log(\cdot)$ factors in the accuracy parameters $\delta,\epsilon$.} $O^*(B)$. This is the dominant model in the theoretical computer science and symbolic computation communities. 

One important point is that even though arithmetic is performed exactly in this model, any algorithm for computing eigenvalues must at some point work with  approximations of them since they need not be rational. The accuracy of these approximations is a function of the maximum bit length $B$ that is used by the algorithm, which we will refer to as bits of precision. \\
\item [(2)] {\bf Finite Precision Arithmetic.} Numbers are stored and manipulated
approximately up to some fixed machine precision $\emach=\emach(n,\delta)>0$. This means each number
$x\in\C$ is stored as $\fl(x)=(1+\delta)x$ for some adversarial multiplicative error $|\delta|\le \emach$, and
each arithmetic operation $\circ\in \{+,-,\times,\div\}$ is guaranteed to yield
an output satisfying 
$$ \fl(x\circ y) = (x\circ y)(1+\delta)\quad |\delta|\le \emach.$$ 
Thus, the outcomes of all operations are (adversarially) noisy due to
roundoff, but the bit lengths of numbers remain fixed at $b:=\log(1/\emach)$. The bit
complexity of the algorithm is the number of arithmetic operations times $O^*(b)$. This is the dominant model in the numerical analysis community, and closely models floating point arithmetic.
\end{enumerate}
In this article we focus mainly on models (1) and (2) above, and occasionally mention exact arithmetic complexity results in passing. A key difference between the two models is that in (1) the algorithm has complete control over how and when to round numbers (which is only necessary when they become too large, and typically leads to {\em additive} errors), whereas {\em all} arithmetic operations incur an adversarial multiplicative rounding error in (2). The special case of rational arithmetic in which all numbers are constrained to have the same denominator (typically a fixed power of $2$ or $10$) is called {\em fixed precision arithmetic}.

It is impossible to ``exactly'' compute eigenvalues and eigenvectors in general when $n\ge 5$ due to the Abel-Ruffini theorem, so they can only be computed approximately\footnote{that is, in the models of computation above, which require storing numbers explicitly. If the output is allowed to be an implicit representation of the eigenpairs in e.g. a field extension, then an exact representation via minimal polynomials is possible; see Remark \ref{rem:symbolic}}. We will consider two notions of approximation:
\begin{enumerate}
    \item [(1)] {\bf Forward Error.} Given an input $A$ and desired accuracy parameter $\epsilon>0$, compute a diagonal $\ax{D}$ and invertible, normalized $\ax{V}$ such that 
    \begin{equation}\label{eqn:forwarddiag}\|\ax{D}-D\|\le \epsilon \|D\|, \|\ax{V}-V\|\le \epsilon \|V\|,\textrm{ and }\|\ax{V}^{-1}-V^{-1}\|\le \epsilon \|V^{-1}\|\end{equation}
    for some exact diagonalization $A=VDV^{-1}$. Here we are asking for an approximation of the exact output corresponding to the input $A$, i.e., the approximation occurs in the output space. This is the standard notion of approximation in theoretical computer science, and is suitable for mathematical applications in which the exact input is meaningful and the output depends discontinuously on the input (e.g., computing eigenvalue multiplicities in representation theory). It is natural to ask for an approximation of $V^{-1}$ as most applications of diagonalization require being able to apply the computed similarity.
    \item [(2)]{\bf Backward Error.} Given an input $A$ and accuracy parameter $\delta>0$, compute $\ax{D}$ and normalized $\ax{V}$ such that
    \begin{equation}\label{eqn:backwarddiag}\|A-\ax{V}\ax{D}\ax{V}^{-1}\|\le \delta\|A\|,\end{equation}
    i.e., compute the exact diagonalization $\ax{A}=\ax{V}\ax{D}\ax{V}^{-1}$ of a nearby matrix $\ax{A}$ satisfying $\|\ax{A}-A\|\le \delta\|A\|$. Note that the approximation occurs in the input space as opposed to the output space. This is the standard notion in numerical analysis and scientific computing where the problem is assumed to be well-posed or well-conditioned, i.e., the output varies continuously with the input. It is further desirable to have a bound on $\|\ax{V}\|\|\ax{V}^{-1}\|$ so that the computed similarity can be stored and applied in finite precision.
\end{enumerate}

The above taxonomy leads to four different versions of the diagonalization problem.  The most well-studied and interesting combinations are (rational arithmetic, forward error) and (finite arithmetic, backward error); the other combinations are subsumed by these in a way which we will make precise in Remarks \ref{rem1} and \ref{rem2}.  We report below on the history and best currently known complexity bounds in these settings, and list several open questions which we hope will interest researchers from all of the communities listed above.

\paragraph{Spectral Stability.} The organizing principle of this survey is that any analysis of an algorithm for diagonalizing a general (not necessarily normal) square matrix must include a way to ensure {\em spectral stability} of the matrices that arise during its execution; by spectral stability we mean that perturbing the entries of a matrix leads to commensurately small perturbations of its eigenvalues and eigenvectors. Two examples of matrices which are spectrally {\em unstable} are \begin{enumerate}
    \item [(E1)]
The $n\times n$ tridiagonal Toeplitz matrix $S_n$ with all $1/4$s on the subdiagonal, all $1$s on the superdiagonal, and zeros elsewhere. \item [(E2)] The $2n+1\times 2n+1$ symmetric tridiagonal matrix $T_n$ with diagonal
$$T_n(k,k)= |n+1-k|\quad k=1,2\ldots,2n+1,$$ and all $1$'s on the superdiagonal and subdiagonal. 
\end{enumerate} 
Both matrices have distinct real eigenvalues. It turns out (see e.g. the excellent book \cite{trefethen2005spectra} for details) that a small perturbation $S_n+E$ of $S_n$ with can change the eigenvalues of $S_n$ by up to $2^n\|E\|$ in the worst case; the culprit is that the eigenvectors of $S_n$ are very far from orthogonal (with condition number $2^n$), which leads to an unstable dependence of the spectrum on the matrix entries. In the example (E2), a perturbation of $T_n$ of size $\|E\|\approx 2^{-n}$ can change the eigenvectors of $T_n$ by $\Omega(1)$. The issue is that the eigenvalue gaps of $T_n$ are exponentially small in $n$ (see \cite{parlett1992minimum,wilkinson1988algebraic}), and  such a perturbation can collapse two nearby eigenvalues and then separate them in a way which produces to a macroscopically different basis for the $2-$dimensional invariant subspace in question, which is problematic for the forward error formulation (1).

The potential appearance of matrices such as $S_n$ and $T_n$ presents a challenge in proving convergence of any algorithm for solving the diagonalization problem. The issue of spectral stability is handled using different tools in different settings; in the (rational arithmetic, forward error) setting, the tools used are algebraic and of a diophantine nature, whereas in the (finite arithmetic, backward error) setting the key input comes from random matrix theory.

\section{Rational Arithmetic, Forward Error}

\newcommand{\Z}{\mathbb{Z}}
\newcommand{\mindgap}{\mathrm{mindgap}}
In this section we will assume that the input is given as an integer matrix and that the bit length of the largest integer is at most $a$; we denote the set of such matrices by $\Z^{n\times n}\bit{a}$. There is no loss of generality here as any $n\times n$ matrix with complex rational entries can be reduced to a $2n\times 2n$ rational matrix by replacing each entry $a+ib$ by a $2\times 2$ matrix $$\begin{bmatrix} a & b\\ -b &a\end{bmatrix},$$  and a forward approximate diagonalization of the former can be read off from a forward approximate diagonalization of the latter. Any rational matrix can then be made integer by clearing denominators; the blowup in the denominators can vary substantially depending on the matrix, and this complexity is reflected in the bit length $a$. 

The goal is now to prove worst case complexity bounds for the forward error diagonalization problem which depend on $n,a,$ and the desired forward error $\epsilon$. The first published result for solving the forward error approximate diagonalization problem is due to Cai \cite{cai1994computing}, who showed more generally that the Jordan Normal Form of an arbitrary $A\in\Z^{n\times n}\bit{a}$  can be computed in  $\poly(n,a,\log(1/\epsilon))$ bit operations (where the degree of the polynomial is not specified, but is seen to be at least twelve). The only other rigorous result we are aware of in this setting is the following recent improvement, where $\omega$ denotes the exponent of matrix multiplication.
\begin{theorem}[\cite{dey2023bit}]\label{thm:jnf}
    There is a randomized algorithm which on input $A\in\Z^{n\times n}\bit{a}$ and $\epsilon>0$ computes a forward approximate diagonalization of $A$ in the sense of \eqref{eqn:forwarddiag} in at most $O^*(n^{\omega+3}a+n^4a^2+n^\omega \log(1/\epsilon))$ bit operations.
\end{theorem}
The above theorem also computes a forward approximate Jordan Normal Form of $A$, which is a more demanding task than diagonalization. Curiously, we are not aware of any results that solve the forward error diagonalization problem without computing the Jordan Normal Form.

The dominant term in the complexity bound of Theorem \ref{thm:jnf} is $O^*(n^{\omega+3}a)$, which corresponds roughly to a constant number of matrix multiplications and inversions carried out on matrices with rational entries of bit length at most $B=O^*(n^3a)$. There are two parameters related to spectral stability which dictate this amount of precision. The first is the {\em minimum gap} between {\em distinct} eigenvalues of $A$, which we denote
\begin{equation}\label{eqn:mindgapdef}
    \mindgap(A):=\min_{\lambda_i\neq\lambda_j}|\lambda_i-\lambda_j|,
\end{equation}
 with the convention that $\mindgap(I)=1$ for the identity matrix. If $\mindgap(A)\ll 2^{-B}$, then storing numbers with $B$ bits of precision is inadequate to distinguish the nearby eigenvalues of $A$ and it is unclear how to obtain correct eigenspaces without doing so. A key ingredient in the proof of Theorem \ref{thm:jnf} is the estimate
 \begin{equation}\label{eqn:est1forward}
     \mindgap(A)\ge 2^{-O^*(an^2)}\quad\textrm{for all}\quad A\in\Z^{n\times n}\bit{a}.
 \end{equation} 
 This follows from two well-known facts: the coefficients of $\det(zI-A)$ are bounded by $O^*(an)$ (see e.g. \cite{grotschel2012geometric}), and the distinct zeros of a univariate polynomial with integer coefficients of magnitude at most $h$ must have distance at least $O^*(h^{-n})$ (\cite{mahler1964inequality}).
 
 The second important parameter is the {\em eigenvector condition number} of $A$, denoted $\kappa_V(A)$, which is defined for diagonalizable $A$ as:
\begin{equation}\label{eqn:kappavdef}
    \kappa_V(A):=\inf_{V} \|V\|\|V^{-1}\|,
\end{equation}
where the infimum is over all invertible $V$ such that $A=VDV^{-1}$ for some diagonal $D$. Note that $\kappa_V(A)=1$ if and only if $A$ is normal, and by the Bauer-Fike theorem, a small value of $\kappa_V(A)$ implies that the spectrum of $A$ is stable under perturbations of its entries. On the other hand, if $\kappa_V(A)\gg 2^B$ then it is impossible to accurately write down both $V$ and $V^{-1}$ using $B$ bits of precision, so $\log(\kappa_V(A))$ is a natural lower bound on $B$. The second key estimate in the proof of Theorem \ref{thm:jnf} is  
\begin{equation}\label{eqn:est2forward}
    \kappa_V(A)\le 2^{O^*(an^3)}\quad\textrm{ for every diagonalizable\footnote{in fact, the same estimate holds for the similarity in the Jordan Normal Form)}}\quad A\in\Z^{n\times n}\bit{a}.
\end{equation} This bound leverages an algebraic result of Geisbrecht and Storjohann \cite{giesbrecht2002computing} on the size of integers appearing in the Frobenius canonical form of an integer matrix, as well as an analytic result of Batenkov \cite{batenkov2012norm} on the conditioning of the eigenvectors of companion matrices in terms of their minimum distinct eigenvalue gaps.

 We briefly describe the algorithm of Theorem \ref{thm:jnf}, which combines tools from the symbolic and numerical computation communities. It first uses the $O(n^5a)$ algorithm of Geisbrecht and Storjohann \cite{giesbrecht2002computing} to {\em exactly} compute the Frobenius Canonical Form $A=UCU^{-1}$ where $C$ is a direct sum of companion matrices and both $U$ and $C$ have controlled bit length. It then {\em approximately} computes the eigenvalues of each companion matrix using the root finding algorithm of Pan \cite{pan2002univariate}, which then yields an approximate diagonalization of $C$ based on explicit formulas for the eigenvectors of companion matrices in terms of their eigenvalues \cite{brand1964companion}. A key step is determining the accuracy to which the roots must be computed, which depends on $\mindgap(A)$ and the condition number $\kappa_V(A)$, both of which can be bounded {\em a priori} via \eqref{eqn:est1forward} and \eqref{eqn:est2forward} above. Combining the approximate diagonalization of $C$ with the similarity $U$ yields an approximate diagonalization of $A$.

\begin{remark}[Symbolic Representations]\label{rem:symbolic}
In the symbolic computation community, the works \cite{kaltofen1986fast, ozello1987calcul, gil1992computation, giesbrecht1995nearly, roch1996fast,li1997determining} gave polynomial {\em exact arithmetic} complexity\footnote{The works \cite{ozello1987calcul, gil1992computation} derived bit complexity bounds for certain special cases of input matrices, but not in general.} bounds for computing the ``rational Jordan form'' of a matrix $A=VJV^{-1}$ over any field; the idea of using the Frobenius Canonical Form as an intermediate step, which Theorem \ref{thm:jnf} is based on, was introduced in these works.  Roughly speaking, the rational Jordan form involves a symbolic representation of the matrices $J$ and $V$ where the eigenvalues are represented in terms of their minimal polynomials over the field. This includes the diagonalization problem over $\C$ as a special case, but the symbolic representation does not always allow one to compute the inverse of $V$ efficiently. A key step in the algorithm of \cite{cai1994computing} was to efficiently extract a numerical representation of the Jordan Normal Form from such a symbolic representation of $V$ and $J$. \end{remark}

 \paragraph{Open Questions.} There are several computational and purely mathematical questions suggested by this line of work.
 \begin{enumerate}
    \item [(Q1)] Can the complexity bound of Theorem \ref{thm:jnf} be improved? A bottleneck to improving the bound beyond $O^*(n^5a)$ is use of \cite{giesbrecht2002computing}, so going beyond this barrier would require a new approach. An ambitious goal might be to obtain a complexity in the ballpark of $O^*(n^{\omega+1}a)$ operations, roughly matching the best known complexity for approximately computing just the eigenvalues (without eigenvectors) of a matrix $A\in\Z^{n\times n}\bit{a}$ \cite{pan1999complexity}.
    \item [(Q2)]What is the complexity of approximate diagonalization when the input $A$ is assumed to be Hermitian? In this case we know that $\kappa_V(A)=1$. Hermitian matrices are perhaps the most common ones in applications, so this special case is well-motivated.
     \item [(Q3)] Can the eigenvalue gap estimate \eqref{eqn:est1forward} be substantially improved if $A$ is restricted to be Hermitian? It was recently shown \cite{abrams2022eigenvalue} that the lower bound in \eqref{eqn:est1forward} is sharp up to a constant factor in the exponent for general $A$, and this bound is realized for certain nonsymmetric matrices. On the other hand, the best known upper bound on the gap for a {Hermitian} matrix is $2^{-O^*(an)}$ (given by Example (E2) \cite{parlett1992minimum, wilkinson1988algebraic}), which is much larger. This is relevant to Question (2) above.
    \item [(Q4)] What is the optimal upper bound in estimate \eqref{eqn:est2forward}? Improvements could be relevant to Question (1) above.
    \item [(Q5)] Is approximate diagonalization strictly easier than approximately computing the Jordan Normal Form, or are there efficient reductions between the two problems? 
 \end{enumerate}

\section{Finite Arithmetic, Backward Error}
In this section we consider the backward error formulation \eqref{eqn:backwarddiag} of the diagonalization problem in  finite arithmetic.  The new features in this setting are: (i) The model of computation can introduce adversarial rounding errors in every step of the algorithm. This is challenging to analyze unless all matrices appearing during the execution of the algorithm are guaranteed to have eigenvalues and eigenvectors which are {stable} under adversarial perturbations of the matrix entries.  (ii) The notion of backward error gives us leeway to perturb the input slightly and absorb the perturbation into the backward error. 

These two features turn out to complement each other very well. The punch line is that (ii) can be used to guarantee the very spectral stability properties which are needed to overcome (i). We now describe two different instantiations of this phenomenon.

The first algorithm with rigorous guarantees in this setting was the fairly recent work \cite{abbcs} in the real number computation community, which gave a randomized worst-case running time of $O(n^9/\delta^2)$ arithmetic operations with $b=O(\log (n/\delta))$ bits of precision, where $\delta>0$ is the desired backward error.  The algorithm is based on the homotopy continuation method, which slowly deforms an ``easy'' instance of the diagonalization problem (e.g., a diagonal matrix $A_0$) into a given input $A_1=A$ via a homotopy $$A_t:=(1-t)A_0+tA_1, t\in [0,1],$$ while keeping track of a forward approximate diagonalization of $A_t$, eventually resulting in a forward approximate diagonalization of $A$ . The speed at which the time parameter $t\in [0,1]$ can be evolved depends on the Euclidean distance of $A_t$ to the set of matrices with a multiple eigenvalue (the set of ``ill-posed'' instances) --- this distance is the (geometric) notion of spectral stability used by the algorithm. 

The key insight of \cite{abbcs} (inspired by the smoothed analysis framework of \cite{spielman2004smoothed}) is to show that that if we replace $A_1$ by $A_1+\delta G$  where $G$ is a complex Gaussian matrix normalized to have $\|G\|= \|A\|$, then with high probability the entire homotopy path is well-separated  from the set of ill-posed instances, and one can perform the homotopy continuation efficiently. It turns out that a sufficiently good forward approximate diagonalization of $A_1+\delta G$ is a roughly $\delta-$backward error diagonalization of $A_1$, resulting in the stated bound.  The proof of this  probabilistic result uses techniques from geometric probability, such as Gaussian integration on manifolds.

The bound of \cite{abbcs} was improved in \cite{banks2020pseudospectral}, who showed the following, which remains the best known complexity bound for backward error diagonalization. Apart from the smaller exponent of $n$, an important feature of this theorem is its logarithmic (rather than polynomial) dependence on $\delta$; however, the amount of precision used is also larger ($\log^4(n/\delta)\log(n)$ as opposed to simply $\log(n/\delta)$).
\newcommand{\MM}{\mathrm{MM}}
\begin{theorem}[\cite{banks2020pseudospectral}] \label{thm:bkwd} There is a randomized algorithm which on input any matrix $A\in \C^{n\times n}$ and $\delta>0$ outputs a diagonal $D$ and invertible $V$ such that 
$$ \quad \|A-VDV^{-1}\|\le \delta\|A\| \quad\mathrm{and}\quad \|V\|\|V^{-1}\| \le O\left(n^{2.5}/\delta\right)$$
in 
$$O\left(T_\MM(n)\log^2\frac{n}{\delta}\right)$$
arithmetic operations in finite arithmetic with $$O(\log^4(n/\delta)\log n)$$ bits of precision, with probability at least $1-14/n$. Here $T_\MM(n)$ refers to the running time of a numerically stable matrix multiplication algorithm, known to satisfy $T_\MM(n)=O(n^{\omega+\alpha})$ for every $\alpha>0$. 
\end{theorem}

\newcommand{\gap}{\mathrm{mingap}}
The algorithm in Theorem \ref{thm:bkwd} is a black box reduction from the diagonalization problem to $O(\log^2(n/\delta))$ matrix multiplications. It is based on the {\em spectral bisection} method developed by Beavers and Denman in the numerical analysis community in the 1970's \cite{beavers1974new}. That method is based on the complex-analytic fact that the function 
\begin{equation}\label{eqn:signfunc} g(A):=(A+A^{-1})/2\end{equation}
maps all of the eigenvalues of $A$  in the open right (resp. left) complex half plane closer to $+1$ (resp. $-1$) in a certain sense; iterating this map rapidly computes an approximation of the {\em matrix sign function} (see \cite{banks2020pseudospectral}[Section 1.2]) of $A$, provided the eigenvalues are initially well-separated from the imaginary axis. It is shown in \cite{demmel2007fast, banks2020pseudospectral} that the backward approximate diagonalization problem can be reduced to computing the sign function in finite arithmetic via a divide and conquer algorithm. While it is easy to analyze the iteration \eqref{eqn:signfunc} in exact arithmetic, it becomes challenging in finite arithmetic since the iterates no longer commute and could be spectrally unstable \cite{byers1986numerical,byers1997matrix}. 

A key contribution of \cite{banks2020pseudospectral} is a proof that if we randomly perturb any $A\in\C^{n\times n}$ by adding an i.i.d. complex Gaussian matrix $A+\delta G$  with $\|G\|= \|A\|$ as in \cite{abbcs}, then with high probability we have
\begin{equation}\label{eqn:shattering}\kappa_V(A+\delta G)\le \poly(n/\delta)\quad and\quad \gap(A+\delta G)\ge \poly(\delta/n),\end{equation}
where $\gap(M)$ denotes the minimum gap between two (not necessarily distinct) eigenvalues of $M$. It can be shown that these properties imply stability of the eigendecomposition under perturbations and are moreover preserved by the iteration \eqref{eqn:signfunc},  enabling its analysis in finite arithmetic and a proof of Theorem \ref{thm:bkwd}.  

The idea of proving a bound like \eqref{eqn:shattering} originates in a conjecture of E. B. Davies \cite{davies2007approximate}, which posits that every matrix $\|A\|$ is $\delta\|A\|$-close in the operator norm to a matrix $A'$ with 
\begin{equation}\label{eqn:davies} \kappa_V(A')\le c_n/\delta,\end{equation} where $c_n$ is a function only of $n$. The conceptual meaning of this conjecture is that every matrix is close to a matrix which is spectrally stable. This conjecture was proven with $c_n=4n^{3/2}$ in \cite{banks2019gaussian} using techniques from nonasymptotic random matrix theory as well as the theory of pseudospectra developed in numerical analysis, which was the first step towards establishing \eqref{eqn:shattering}.

We have now seen two substantially different algorithms for the backward approximate diagonalization problem with one commonality: the use of a complex Gaussian perturbation as a preprocessing step. In \cite{abbcs} this led to a certain geometric notion of spectral stability, whereas in \cite{banks2020pseudospectral} it led to bounds on $\gap$ and $\kappa_V$, a seemingly different notion of spectral stability. It was later realized that these two notions of spectral stability are actually closely related, and each one can be used to derive the other --- see \cite[Appendix D]{banks2020pseudospectral} for details. At a high level this connection exists because there are two closely related ways to define the condition number of a problem (see \cite{demmel1987condition}): as the distance to an ill-posed set of instances, or as a measure of the stability of the output as a function of small perturbations of the input.

\begin{remark}[Forward Error in Finite Arithmetic]\label{rem1} It can be shown \cite[Proposition 1.1]{banks2020pseudospectral} that if $A$ has distinct eigenvalues, then a $\delta$-backward approximate diagonalization of $A$ is an $\epsilon$-forward approximate diagonalization of $A$ with $$\epsilon \le \frac{\poly(n)\kappa_V(A)}{\gap(A)}\times \delta.$$ Combining this with Theorem \ref{thm:bkwd} yields an algorithm for the forward approximate eigenproblem in finite arithmetic with logarithmic dependence on $1/\epsilon, \kappa_V(A),$ and $1/\gap(A)$. This reduction can be extended to matrices with multiple eigenvalues if one uses condition numbers for invariant subspaces of $A$ rather than eigenvectors; the reader may consult any numerical linear algebra textbook such as \cite{demmel1997applied} for details.
\end{remark}
\begin{remark}[Backward Error in Rational Arithmetic]\label{rem2} There has so far not been any benefit in considering backward error algorithms in the rational arithmetic model. The heuristic reason is that exact basic linear algebra operations such as inversion and QR factorization require at least $n$ bits of precision in the rational arithmetic model, since inverting an integer matrix can blow up the bit length of its entries by a factor of $n$ in the worst case \cite{grotschel2012geometric}. All known approaches to diagonalization require performing one of these basic operations, and consequently must have bit complexity at least $\Omega(n^{\omega +1})$, which is already much higher than the the bit complexity guaranteed by Theorem \ref{thm:bkwd} in the finite arithmetic model. 
\end{remark}

\paragraph{Open Questions.} While the nearly matrix multiplication time bound of Theorem \ref{thm:bkwd} seems quite satisfactory, much remains to be done in this area. We conclude by listing four questions which seem quite mysterious at the moment.
\begin{enumerate}
    \item [(Q6)] Can the bits of precision used by Theorem \ref{thm:bkwd} be reduced to $O(\log(n/\delta))$? This would be more in line with the notion of ``numerical stability'' in numerical analysis, and match the precision required in the Hermitian case \cite{parlett1998symmetric} (see  also \cite{banks2022global}). Such a low precision is achieved by \cite{abbcs}, which however has a large running time. 
    \item [(Q7)] Is there a {\em deterministic} efficient algorithm for the backward error diagonalization problem in finite arithmetic? Both of the algorithms presented in this section use a random perturbation. Is there a deterministic algorithm for finding a perturbation guaranteeing the spectral stability bounds \eqref{eqn:shattering}?
    \item [(Q8)] What is the optimal dependence $c_n$ in \eqref{eqn:davies}? In particular, can the dependence on $n$ be removed altogether? This is closely related to an old conjecture of Davidson, Herrero, and Salinas \cite{davidson1989quasidiagonal}in operator theory \cite[Problem I.17]{davidson2001local}.
    \item [(Q9)] Can spectral the stability guarantees \eqref{eqn:shattering} be shown for a {\em sparse} random perturbation (instead of the dense complex Gaussian matrix above)? This would improve the efficiency of this step in practice and may widen the scope of its applicability.
\end{enumerate}

\bibliographystyle{alpha}
\newcommand{\etalchar}[1]{$^{#1}$}

\end{document}